\documentclass[twocolumn,aps,prl,superscriptaddress]{revtex4}
\usepackage[utf8]{inputenc}
\usepackage{epsf,graphicx}
\usepackage{multirow}
\usepackage{amsmath,gensymb,amssymb}
\usepackage{lipsum}
\usepackage{dcolumn}
\usepackage{bm}
\usepackage{hyperref}
\usepackage{natbib}
\usepackage{color}

\renewcommand{\vec}{\mathbf }

\begin{document}

\title{Simulation of non-stationary anisotropic capillary wave turbulence within the framework of the kinetic equation}

\author{E.\,A.~Kochurin}
\email{kochurin@iep.uran.ru}
\affiliation{Skolkovo Institute of Science and Technology, 121205, Moscow, Russia}
\affiliation{Institute of Electrophysics, Ural Division, Russian Academy of Sciences, Yekaterinburg, 620016 Russia}

\author{A.\,O.~Korotkevich}
\email{kochurin@iep.uran.ru}
\affiliation{Skolkovo Institute of Science and Technology, 121205, Moscow, Russia}
\affiliation{Landau Institute for Theoretical Physics, RAS, Chernogolovka, 142432, Moscow
	region, Russia }

\begin{abstract}In the present work, numerical simulation of the evolution of capillary turbulence was performed on the basis of the wave kinetic equation taking into account energy dissipation at small scales. For an isotropic initial condition, the simulation results reproduce the Zakharov--Filonenko power-law turbulence spectrum for the wave action, \(n_k \sim k^{-17/4}\). The main result of the work is the calculation of the evolution for a substantially anisotropic initial condition.
The numerical data indicate that, during the evolution of an anisotropic wave-action distribution, a factorized spectrum \(n(k,\phi) \sim k^{-17/4}F(\phi)\) is formed in polar coordinates \((k,\phi)\), where \(F(\phi)\) is a function of the angle in Fourier space.
In the regime of a steady power-law distribution in \(k\), the calculated collision integral vanishes with high accuracy in the inertial interval of wavenumbers (the region of the power-law spectrum). Thus, the work provides evidence for the realization of a new strongly anisotropic turbulence spectrum, similar to Kolmogorov--Zakharov spectra.
\end{abstract}

\maketitle

\textbf{Introduction.}
The theory of weak wave turbulence, which describes statistically weakly nonlinear wave interactions in various media, was developed in the works of V.~E.~Zakharov and coauthors~\cite{ZakharovPhD,ZF67,Zakh67,ZZ82}. This theory is based on the wave kinetic equation, which describes the evolution of a universal physical quantity -- the wave action $n(\vec{k};t)$, which has the meaning of the spectral density of the number of waves. Exact stationary solutions of the kinetic equations corresponding to nonzero fluxes of conserved quantities are called Kolmogorov--Zakharov turbulence spectra, by analogy with classical turbulence~\cite{ZLF1992,Nazarenko2011}. These exact solutions correspond to a special, distinguished case in which the collision integral $St$ describing nonlinear wave interactions vanishes: $St=0$. From a physical point of view, this situation corresponds to constant fluxes of conserved quantities (energy and wave action) along scales in the inertial interval of wavenumbers, where viscosity and pumping do not have a significant effect.

One of the most important applications of weak turbulence theory is the description of wave turbulence on the surface of a three-dimensional fluid, in particular, water waves. For small-scale capillary waves, for which gravity effects are negligible, the theory predicts a power-law wave-action spectrum, \(n(\mathbf{k})\sim k^{-17/4}\); see~\cite{ZakharovPhD,ZF67}. This exact power-law solution, known as the Zakharov--Filonenko spectrum, is valid for stationary isotropic capillary turbulence. Detailed experimental studies~\cite{exp1,exp2,exp3,exp4,exp5,exp6,exp7,exp8,exp9,exp10} and numerical studies~\cite{PZ,Push1,Push2,DNS1,DNS2} reproduce the Zakharov--Filonenko spectrum with high accuracy. Thus, to date, isotropic capillary turbulence is one of the most well-studied types of wave turbulence. The situation changes substantially when an anisotropic regime of wave propagation is considered. Furthermore, it is precisely the anisotropic regime that better corresponds to the physics of the process, since in real systems surface waves are excited by wind forcing, which is, as a rule, anisotropic.
In the weak-anisotropy approximation, when the spectrum is modulated with respect to the azimuthal angle \(\phi\) in Fourier space, the Kats--Kontorovich spectrum is valid~\cite{KK1,KK2}. In the opposite limiting case, when all waves propagate exclusively in one direction, the turbulence spectrum is generated by resonant five-wave interactions and does not coincide with the Zakharov--Filonenko spectrum obtained under the assumption of the dominant influence of three-wave interactions~\cite{1D1,1D2,1D3}. The intermediate case, in which the anisotropy of wave propagation is arbitrary, remains virtually unexplored.

The aim of this study is to numerically investigate the anisotropic regime of capillary turbulence within the framework of a kinetic equation. The computational algorithm employed is based on calculating the collision integral along the resonance interaction curve, with an additional account for small-scale energy dissipation. It is shown that nonlinear wave interactions generate a power-law distribution of wave action, $n({\bf k})\sim k^{-17/4}$, not only for an isotropic initial condition, but also in the anisotropic case, when waves predominantly propagate in one direction. It is important that for the detected strongly anisotropic regime the collision integral turns to zero with high accuracy, $St=0$, which indicates the stationarity of the spectrum.

\textbf{Kinetic equation.}
In this work, we study the wave turbulence of capillary waves on the free two-dimensional surface of a three-dimensional fluid. Specifically, if we consider deep water under standard conditions, we are examining waves on scales smaller than 5 mm, for which the influence of gravity can be neglected. The problem of capillary turbulence was first addressed by V.~E.~Zakharov \cite{ZakharovPhD} and, subsequently, in his joint work with N.~N.~Filonenko \cite{ZF67}. In these studies, the kinetic equation for capillary waves was derived, and an exact stationary solution was obtained, corresponding to a direct turbulent energy cascade (from large scales to small scales). Since then, this solution has been the subject of numerous theoretical and experimental investigations; see, for example, the review \cite{Galt}. The model of capillary wave turbulence is based on the fundamental weakly nonlinear equations for the potential flow of an ideal, incompressible fluid with a free surface. It is assumed that, in the state of rest, the fluid surface coincides with the $XY$-plane in a Cartesian coordinate system, while the fluid itself occupies the lower half-space $z \le 0$. The fluid velocity at any given point is defined by $\mathbf{v} = \nabla\Phi$, where $\Phi = \Phi(x, y, z; t)$ is the velocity potential. In this paper, we do not present the original dynamical equations, as the formalism of weak turbulence theory and the derivation of the kinetic equation can be found in well-known works such as \cite{ZakharovPhD, ZF67, ZLF1992, Nazarenko2011, Pan_PHD, Galt}. The primary physical variables in the problem are the shape of the fluid surface, $z=\eta(x,y;t)$, and the value of the velocity potential at the fluid surface, $\psi(x,y;t)=\Phi(x,y,\eta(x,y;t);t)$. By employing an expansion in terms of a small parameter (the mean steepness of the fluid surface, $\mu=\sqrt{\langle|\nabla \eta|^2\rangle}\ll1$), the governing equations and the boundary conditions at the surface are reduced to a pair of Hamiltonian equations~\cite{Zakh67} for $\eta$ and $\psi$ directly. Next, complex normal (canonical) variables are introduced:
\begin{equation}\label{avar}
	a_{\vec k}=\sqrt{\frac{\omega_k}{2k}}\eta_{\vec k}+i\sqrt{\frac{k}{2\omega_k}}\psi_{\vec k},
	\end{equation}
where $\eta_{\vec k}$ and $\psi_{\vec k}$ are the Fourier images of the functions $\eta$ and $\psi$, respectively (we use the symmetric form of the Fourier transform, following the original works). Representing the wave field in terms of these variables is equivalent to a decomposition into plane waves. The wave vector and frequency in \eqref{avar} are related by the dispersion relation for linear capillary waves:
\begin{equation}\label{disp}
	\omega_k=\sqrt{\sigma' k^{3}},\qquad k=|{\vec k}|=\sqrt{k_x^2+k_y^2},
\end{equation}
where $\sigma' = \sigma/\rho$, $\sigma$ is the surface tension coefficient, and $\rho$ is the fluid density.
For further analysis, it is convenient to switch to dimensionless variables.
We have chosen a characteristic wavenumber $k_0 = 2\pi/\lambda_0$ such that it exceeds the critical value $k_c = \sqrt{g/\sigma'}$ (where $g$ is the acceleration due to gravity) -- the point at which the terms in the dispersion relation $\omega_k = \sqrt{gk + \sigma' k^3}$ for waves on the surface of a deep fluid become comparable: $k_0 > k_c$. The time scale $t_0$ can be determined from the dispersion relation~\eqref{disp}: $t_0 = 2\pi/\omega(k_0)$. For instance, for deep water under standard conditions on the Earth's surface, these characteristic values can be estimated as $\lambda_c \approx 1.7$ cm, $\lambda_0 \approx 1$ cm, and $t_0 \approx 0.05$ s. The quantities $t_0$ and $\lambda_0$ will serve as the units of time and length, respectively.
In these new dimensionless units, the dispersion relation (\ref{disp}) takes the form: $\omega_k = k^{3/2}$.

The primary quantity used in the derivation of the kinetic equation is the pair correlator $n_{\vec k}$ (wave action):
\begin{equation*}
	\langle a_{\vec k}a_{{\vec k}'}^*\rangle=n_{\vec k}\delta({\vec k}-{\vec k}'),
\end{equation*}
where the angle brackets denote an ensemble average, and $\delta({\vec k}-{\vec k}')$ is the Dirac delta function. The most detailed derivation of the kinetic equation for capillary waves can be found in \cite{Pan_PHD}. Here, we omit the intermediate calculations and present only the final result:
\begin{equation}\label{kin}
	n_t=St=I_1+I_2+I_3,
\end{equation}
where $St$ is the collision integral, consisting of three terms:
	\begin{multline*} I_1=-4\pi\iint\left[V_{\vec k \vec k_1 \vec k_2}^2\left(n_{{\vec k}}n_{{\vec k}_2}+n_{{\vec k}}n_{{\vec k}_1}-n_{{\vec k}_1}n_{{\vec k}_2}\right)\right.\\
\left.\delta({\vec k}-{\vec k}_1-{\vec k}_2)\delta(\omega-\omega_1-\omega_2)\right]d{\vec k}_1 d{\vec k}_2,\end{multline*}
	\begin{multline*}I_2=4\pi\iint\left[V_{\vec k_1 \vec k \vec k_2}^2\left(n_{{\vec k}_1} n_{{\vec k}_2}+n_{{\vec k}_1} n_{{\vec k}}-n_{{\vec k}}n_{{\vec k}_2}\right)\right.\\
\left.\delta({\vec k}_1-{\vec k}-{\vec k}_2)\delta(\omega_1-\omega-\omega_2)\right]d{\vec k}_1 d{\vec k}_2,\end{multline*}
	\begin{multline*}I_3=4\pi\iint\left[V_{\vec k_2 \vec k \vec k_1}^2\left(n_{{\vec k}_2} n_{{\vec k}_1}+n_{{\vec k}_2} n_{{\vec k}}-n_{{\vec k}}n_{{\vec k}_1}\right)\right.\\
\left.\delta({\vec k}_2-{\vec k}-{\vec k}_1)\delta(\omega_2-\omega-\omega_1)\right]d{\vec k}_1 d{\vec k}_2.\end{multline*}
Here $V_{\vec k \vec k_1 \vec k_2}$ is the matrix element for the interaction of waves:
	\begin{multline*}
		V_{\vec k \vec k_1 \vec k_2}=\frac{1}{8\pi \sqrt{2}}\left(\omega_{{\vec k}}\omega_{{\vec k}_1}\omega_{{\vec k}_2}\right)^{1/2}\\
\left(\frac{L_{{\vec k}_1,{\vec k}_2}}{(k_1k_2)^{1/2}k}-\frac{L_{{\vec k},-{\vec k}_1}{}}{(kk_1)^{1/2}k_2}-\frac{L_{{\vec k},-{\vec k}_2}{}}{(kk_2)^{1/2}k_1}\right),
	\end{multline*}
where $L_{{\vec k}_1,{\vec k}_2}=({\vec k}_1 \cdot{\vec k}_2)+k_1k_2$. Note that the integrals $I_2$ and $I_3$ are equal to each other due to the symmetry under the exchange of ${\vec k}_1$ and ${\vec k}_2$.

An exact stationary solution to equation~\eqref{kin} for the inertial range (a range of scales free from forcing and dissipation) in the isotropic case was first obtained in the dissertation of V. E. Zakharov~\cite{ZakharovPhD} and subsequently published in~\cite{ZF67}, under the assumption that the collision integral vanishes, i.e., $St = 0$. It can be written in the form
\begin{equation}
\label{KZ}
n({\vec k})=2\pi C_0 P^{1/2}k^{-17/4},
\end{equation}
where $C_0$ is the Kolmogorov-Zakharov constant and $P$ is the energy dissipation rate per unit area. The solution \eqref{KZ} is known as the Zakharov-Filonenko spectrum; it is a special case of the Kolmogorov-Zakharov spectrum for capillary waves \cite{ZLF1992, Nazarenko2011}.
The first analytical estimate of the constant $C_0$ in \eqref{KZ} was obtained in \cite{PZ}, yielding a value of 9.85; this was later corrected in \cite{Pan1} to $C_0 = 6.97$ (a factor of $1/\sqrt{2}$ had been omitted).

We briefly outline the algorithm for solving the kinetic equation~\eqref{kin}; a detailed description will be provided in future work. The computational code is developed in the spirit of the well-known Resio-Tracy algorithm~\cite{Resio,Resio2} and its modern modifications~\cite{WRT,Geog1,Geog2}, which are used to model large-scale gravity waves.
The algorithm is based on computing the collision integral directly along the resonant interaction curve. Linear interpolation is employed to determine the values of $n({\vec k})$ on this resonant curve. For the zero harmonic, $n({\vec k})$ is set to zero. Time integration is performed using a 4/5-order Runge-Kutta-Fehlberg method with an adaptive time step. The calculation of the term $St$ can be executed using parallel multiprocessor computing. The simulations were conducted over the time interval $t\in[0, 500]$ with a resolution of $N^2=200^2$ and a wavenumber integration step of $\Delta k=0.5$. The integration domain in $k$-space was $[-50, 50)\times[-50, 50)$. To incorporate the effect of viscosity, we rewrite equation~\eqref{kin} as follows:
\begin{equation}\label{kin2}
	n_t({\vec k})=St(n)-\gamma_k n({\vec k}),
\end{equation}
where we have introduced a viscosity operator defined by $\gamma_k=\gamma_0(k-k_d)^2$ for $k>k_d$, and $\gamma_k=0$ for $k\leq k_d$. The dissipation coefficient was set to $\gamma_0=1$, and the dissipation scale was chosen as $k_d=40$.

	\begin{figure}[t]
\centering
		\includegraphics[width=0.9\linewidth]{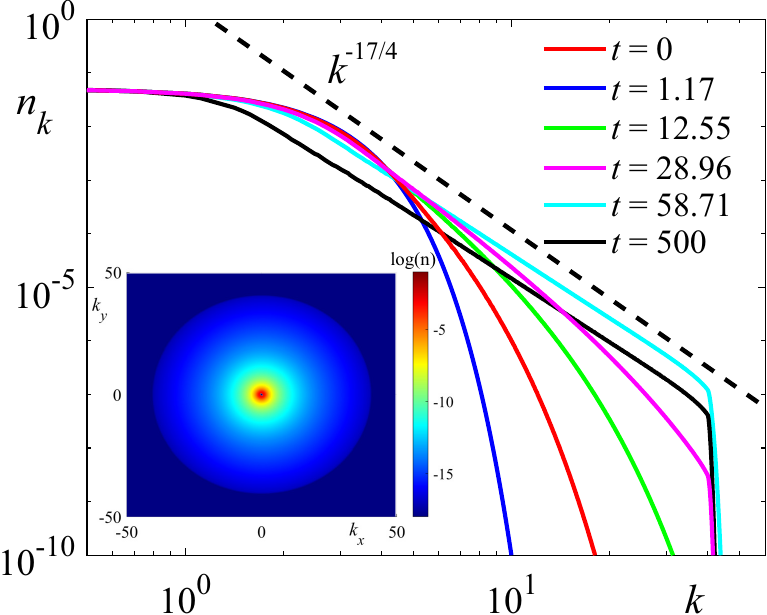}
		\caption{Fig. 1 (Color online). The quantity $n({\vec k})$ averaged over the angle in Fourier space; the black dashed line corresponds to the Zakharov-Filonenko spectrum~\eqref{KZ}. The inset shows the natural logarithm $\log[n({\vec k})]$ of the wave action at time $t=500$. The results were obtained for the initial condition~\eqref{IC1}.}
		\label{fig1}
	\end{figure}

	\begin{figure}[t]
\centering
		\includegraphics[width=0.9\linewidth]{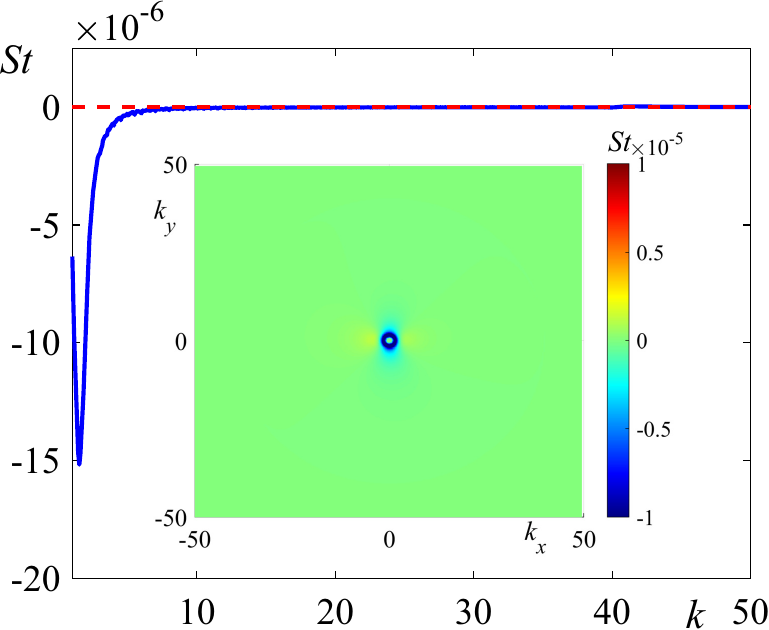}
		\caption{Fig. 2 (Color online). The collision integral $St$ averaged over the angle in Fourier space at time $t=500$ compared with $St=0$. The inset shows the collision integral $St(\vec{k})$ at the same instant in time. The results were obtained for the initial condition (\ref{IC1}).}
		\label{fig2}
	\end{figure}

\textbf{Isotropic turbulence.} Let us first consider an isotropic distribution of wave action:
\begin{equation}\label{IC1}n({\vec k};t=0)=n_0\exp(-[k_x^2+k_y^2]/5),
\end{equation}
where $n_0$, the amplitude of the wave-action perturbation, is set to $n_0=0.05$. For this chosen amplitude $n_0$, the mean initial surface steepness, defined as
\begin{equation}\label{mu}
	\mu=\left(\frac{1}{4\pi^2}\int k^{3/2} n({\vec k})d{\vec k}\right)^{1/2},
\end{equation}
is estimated to be $\mu\approx0.25$; that is, the condition for small surface slopes, $\mu^2\ll 1$, is satisfied. The calculation results for the initial condition (\ref{IC1}) are shown in Fig.~\ref{fig1}. One can observe the transition of the initial Gaussian wave-action distribution into a power-law function by time $t\approx 58.7$. Notably, the exponent is close to the exact value of $-17/4$. Once the power-law distribution is reached, $n({\vec k};t)$ begins to decay in a self-similar manner while maintaining the same exponent. By the end of the simulation interval, $t=500$, the mean surface steepness has decreased to $\mu\approx 0.18$. The wave-action distribution at $t=500$ is shown on a logarithmic scale in the inset of Fig.~\ref{fig1}. It is important that the quantity $n({\vec k})$ is completely isotropic, which indicates that the collision integral in (\ref{kin}) has been calculated correctly.

The solution (\ref{KZ}) is derived under the assumption that the collision integral $St$ is equal to zero (a condition corresponding to the isotropy and stationarity of the turbulence). Consequently, calculating the collision integral in the turbulent regime serves as a useful test for how close the numerical solution of equation (\ref{kin2}) is to the exact solution (\ref{KZ}). Figure \ref{fig2} displays the collision integral, averaged over the angle in $k$-space, at time $t=500$. It can be seen that the value of $St$ remains close to zero across nearly the entire range of wavenumbers. The exception is the region of small wavenumbers, where the initial wave energy is concentrated. In this region, the integral $St$ is negative, which, from a physical point of view, corresponds to the process of energy transfer toward smaller scales. The inset in Fig.~\ref{fig2} shows the spatial distribution of $St(k_x, k_y)$ at $t=500$. It is evident that the collision integral is distributed quite isotropically and rapidly approaches zero within the power-law spectrum region. Thus, for the isotropic initial condition (\ref{IC1}), the simulation results are in good agreement with the Zakharov-Filonenko spectrum (\ref{KZ}).

	\begin{figure}[t]
\centering
		\includegraphics[width=0.9\linewidth]{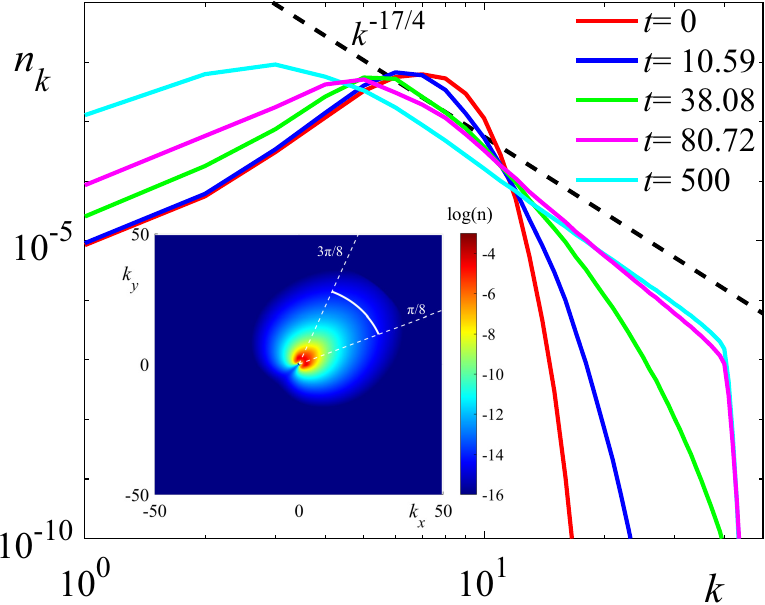}
		\caption{Fig. 3 (Color online). The quantity $n({\vec k})$, averaged over the angle in Fourier space within the range $\phi\in[\pi/8, 3\pi/8]$; the black dashed line corresponds to the Zakharov-Filonenko spectrum (\ref{KZ}). The inset shows the natural logarithm of the wave action, $\log[n({\vec k})]$, at time $t=500$. The results were obtained using the initial condition (\ref{IC2}).}
		\label{fig3}
	\end{figure}

	\begin{figure}[t]
\centering
		\includegraphics[width=0.9\linewidth]{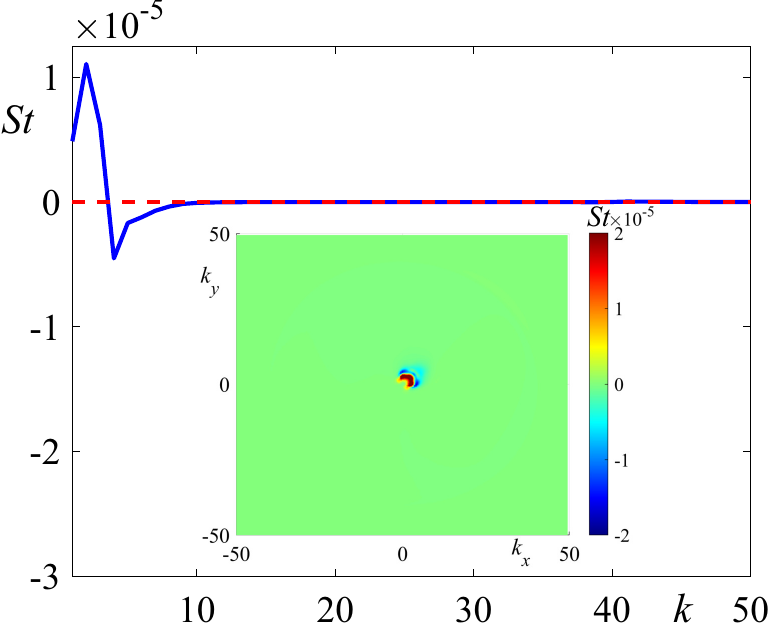}
		\caption{Fig. 4 (Color online). The collision integral $St$ averaged over the angle in Fourier space at time $t=500$ compared with $St=0$. The inset shows the collision integral $St(\vec{k})$ at the same instant in time. The results were obtained for the initial condition (\ref{IC2}).}
		\label{fig4}
	\end{figure}

	\begin{figure}[t]
\centering
		\includegraphics[width=0.9\linewidth]{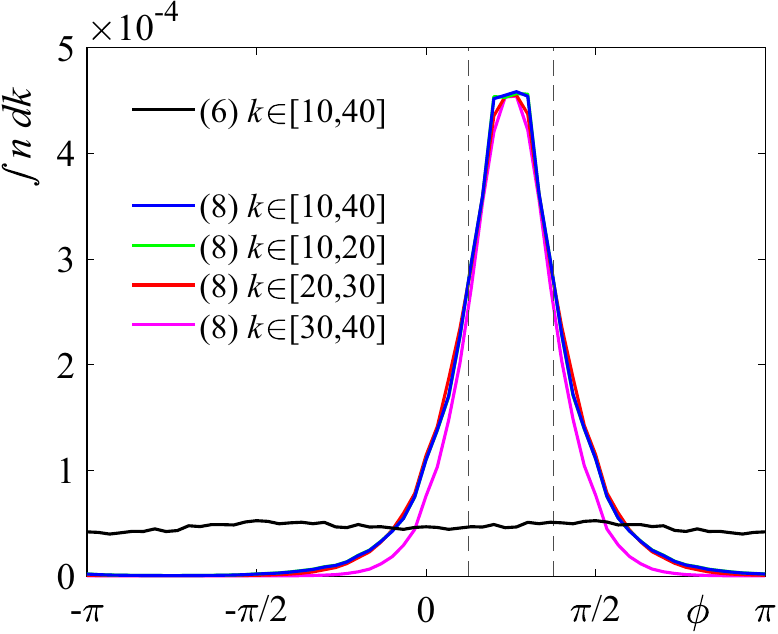}
		\caption{Fig. 5 (Color online). Wave-action integral with respect to $|{\vec k}|$ as a function of the angle in Fourier space, calculated at time $t=500$, for isotropic (black line) and anisotropic (colored lines) initial conditions. The vertical dashed lines correspond to the angles $\pi/8$ and $3\pi/8$.}
		\label{fig5}
	\end{figure}

\textbf{Anisotropic turbulence.}
Let us now consider the evolution of an initially anisotropic distribution of wave action:
\begin{equation}
	\label{IC2}
	n({\vec k};t=0)=n_0\exp(-[(k_x-k_{x_0})^2+(k_y-k_{y_0})^2]/5).
\end{equation}
The distribution \eqref{IC2} also takes the form of a localized function; however, its maximum is not located at the origin, but rather at a specific wave vector ${\vec k}_0=\{k_{x_0}, k_{y_0}\}$. For the second series of calculations, the key parameters in \eqref{IC2} were chosen as follows: $n_0=0.01$ and ${\vec k}_0=\{5, 5\}$. This means that the maximum amplitude of the initial condition, which defines the dominant direction of wave propagation, corresponds to an azimuthal angle of $\phi=\pi/4$. Given these parameters, the mean initial surface steepness is estimated to be $\mu\approx0.28$.

Figure~\ref{fig3} displays the distributions of wave action averaged over the angle $\phi$ in Fourier space within the range $[\pi/8, 3\pi/8]$ at successive time intervals. The choice of the angular sector $\phi \in [\pi/8, 3\pi/8]$ is necessitated by the strong anisotropy of the distribution $n(\vec{k})$.
Overall, the evolution observed in the figure is similar to the result obtained for an isotropic initial condition, as shown in Figure~\ref{fig1}. The quantity $n(\vec{k})$ relatively quickly (by $t \approx 80.7$) transitions to a power-law distribution within the inertial range of wavenumbers, subsequently decaying slowly while maintaining the same power-law exponent. By the end of the simulation period, the mean surface steepness was $\mu \approx 0.22$.
The inset in Figure~\ref{fig3} shows the spatial distribution of wave action at the end of the simulation ($t = 500$). The figure also indicates the angular sector over which the averaging was performed. It is evident that $n(k_x, k_y)$ exhibits significant anisotropy in wavenumber space.
Furthermore, the final result shown in the inset differs substantially from the initial Gaussian form~\eqref{IC2}. It is important to note that once the power law is established, $n(k_x, k_y, t)$ evolves in a self-similar manner: the spatial pattern shown in the inset of Figure~\ref{fig3} remains nearly invariant over time. It appears that this anisotropic wave-action distribution represents a physically distinct, stable state of the system. To test this hypothesis, Fig.~\ref{fig4} shows the angle-averaged collision integral $St$, calculated at time $t=500$. The figure reveals an interesting result: within the range of wavenumbers corresponding to the power-law spectrum, the collision integral vanishes. This is particularly evident in the inset of Fig.~\ref{fig4}. Deviations from zero in the collision integral $St$ are observed only in the large-scale region, which contains the bulk of the initial wave energy. Thus, for anisotropic turbulence, a scenario involving the formation of a stationary, Kolmogorov-type power-law spectrum~\eqref{KZ}, for which the collision integral vanishes, is also realized across a wide range of scales.

In conclusion, we examine the angular dependence of the wave action for the initial conditions \eqref{IC1} and \eqref{IC2}. To do this, we consider the integral of the wave action with respect to the magnitude of the wave vector, $\int_{k_1}^{k_2} n(k, \phi) \, dk$, where the wavenumbers $k_1$ and $k_2$ are chosen such that the integration is performed over a range of wavenumbers corresponding to the power-law spectrum. The calculated angular dependencies are shown in Figure \ref{fig5}. As expected, for the isotropic initial condition \eqref{IC1}, the resulting curve is independent of the angle $\phi$; the minor fluctuations observed are due to the non-uniformity of the angular grid. In contrast, the angular dependence for the anisotropic initial condition \eqref{IC2} exhibits non-trivial behavior, with the integral $\int n(k, \phi) \, dk$ taking the form of a localized function. The figure presents results for four integration intervals: the full power-law spectrum range $[10, 40]$ and the sub-intervals $[10, 20]$, $[20, 30]$, and $[30, 40]$. For the purpose of comparing the angular dependencies, the maxima have been shifted to a single point. It can be seen that the resulting curves are nearly independent of the integration interval: the results essentially overlap, with the exception of the $[30, 40]$ interval. This discrepancy is likely attributable to dissipative effects; although formally absent in this range of scales, they influence the spectrum at high wavenumbers through nonlinear interactions, an effect similar to those observed in \cite{KPZ2019}. Overall, the results obtained reveal a non-trivial fact: the factorization of the solution to the kinetic equation~\eqref{kin} into radial and angular components, namely, $n(k,\phi)=k^{-17/4}F(\phi)$, where $F(\phi)$ is a certain function of the angle.

In this work, we have performed a numerical simulation of capillary turbulence within the framework of the kinetic equation~\eqref{kin2}, accounting for energy dissipation at small scales. The results for an isotropic initial condition~\eqref{IC1} show good agreement with the analytical Zakharov-Filonenko spectrum~\eqref{KZ}: as the system evolves, an isotropic power-law distribution of wave action emerges. For an anisotropic initial condition~\eqref{IC2}, the wave action spectrum evolves in a similar manner. Three-wave nonlinear interactions lead to the formation of a power-law distribution for $n(k)$ with an exponent close to $-17/4$. Once this power-law distribution has developed, the evolution becomes self-similar, so that the shape of \(n(\vec{k};t)\) remains nearly invariant over time. In this self-similar regime, the calculated collision integral $St$ vanishes with high accuracy within the range of the power-law spectrum.
Thus, this work reports the first observation of a strongly anisotropic spectrum of capillary turbulence.
It should be noted that the distribution obtained here is not a special case of the Katz-Kontorovich wave turbulence spectrum~\cite{KK1,KK2}, which is derived under the assumption of weak anisotropy. In future work, we plan to incorporate an energy-pumping term into the kinetic equation~\eqref{kin2} to conduct a detailed analysis of stationary turbulence.

{\bf Funding.}
This work was carried out within the framework of Russian Science Foundation grant No. 25-72-31023.

{\bf Conflict of interest.}
The authors declare no conflict of interest.

{\bf Acknowledgments.}

This work was performed using the computational resources of the Laboratory of Computational Engineering at the Skolkovo Institute of Science and Technology.

\end{document}